\documentclass[12pt,manuscript]{aastex}
\usepackage[dvips]{color}
\begin{document}

\newcommand{\kms}{\mbox{km~s$^{-1}$}}
\newcommand{\s}{\mbox{$''$}}
\newcommand{\mloss}{\mbox{$\dot{M}$}}
\newcommand{\my}{\mbox{$M_{\odot}$~yr$^{-1}$}}
\newcommand{\ls}{\mbox{$L_{\odot}$}}
\newcommand{\um}{\mbox{$\mu$m}}
\newcommand{\ujy}{\mbox{$\mu$Jy}}
\newcommand{\ms}{\mbox{$M_{\odot}$}}
\newcommand\mdot{$\dot{M}  $}

\newcommand{\vexp}{\mbox{$V_{\rm exp}$}}
\newcommand{\vsys}{\mbox{$V_{\rm sys}$}}
\newcommand{\vlsr}{\mbox{$V_{\rm LSR}$}}
\newcommand{\tex}{\mbox{$T_{\rm ex}$}}
\newcommand{\teff}{\mbox{$T_{\rm eff}$}}
\newcommand{\tmb}{\mbox{$T_{\rm mb}$}}
\newcommand{\trot}{\mbox{$T_{\rm rot}$}}
\newcommand{\tkin}{\mbox{$T_{\rm kin}$}}
\newcommand{\dens}{\mbox{$n_{\rm H_2}$}}
\newcommand{\bri}{\mbox{erg\,s$^{-1}$\,cm$^{-2}$\,\AA$^{-1}$\,arcsec$^{-2}$}}
\newcommand{\brib}{\mbox{erg\,s$^{-1}$\,cm$^{-2}$\,arcsec$^{-2}$}}
\newcommand{\flux}{\mbox{erg\,s$^{-1}$\,cm$^{-2}$\,\AA$^{-1}$}}
\newcommand{\ha}{\mbox{H$\alpha$}}

\title{Are Large, Cometary-Shaped Proplyds really (free-floating) EGGs? }
\author{R. Sahai\altaffilmark{1}, R. G{\"u}sten\altaffilmark{2}, M. R. Morris\altaffilmark{3}
}

\altaffiltext{1}{Jet Propulsion Laboratory, MS\,183-900, California
Institute of Technology, Pasadena, CA 91109}

\altaffiltext{2}{Max-Planck-Institut f{\"u}r Radioastronomie, Auf dem H{\"u}gel 69, 53121 Bonn, Germany}

\altaffiltext{3}{Department of Physics and Astronomy, UCLA, Los Angeles, CA 90095-1547}

\email{raghvendra.sahai@jpl.nasa.gov}
\begin{abstract}
We report the detection of strong and compact molecular line emission (in the CO J=3--2,\,4--3,\,6--5,\,7--6, $^{13}$CO J=3--2, 
HCN and HCO$^+$ J=4--3 transitions) from a cometary-shaped object (Carina-frEGG1) in the Carina star-forming region (SFR) previously classified as a photoevaporating protoplanetary disk (proplyd). We derive a molecular mass of 0.35\ms~for Carina-frEGG1, which shows that it is not a proplyd, but belongs to a 
class of free-floating evaporating gas globules (frEGGs) recently found in the Cygnus SFR by Sahai, Morris \& Claussen (2012). Archival Adaptive Optics near-IR (Ks) images show a central hourglass-shaped nebula. The derived source luminosity 
(about $8-18\,\ls$), the hourglass morphology, and the presence of collimated jets seen in HST images, imply the presence of a jet-driving, young, low-mass star deeply embedded in the dust inside Carina-frEGG1. Our results suggest that 
the true nature of many or most such cometary-shaped objects seen in massive SFRs and previously labelled as proplyds has been misunderstood, and that these are really frEGGs.
\end{abstract}
\section{Introduction}
Massive stars have a profound effect on stars that subsequently form in their vicinity, thus significantly influencing the
initial mass function (IMF), the star formation efficiency, and the total extent and mass of star clusters. In massive star-forming regions (SFRs),
the ionizing radiation from OB stars can produce photo-evaporating protoplanetary disks or
{\it proplyds}, the most famous of which lie in the
Orion Nebula (e.g., O'Dell, Wen, \& Hu 1993; Ricci, Robberto,
\& Soderblom 2008). This radiation
can also progressively ionize dense knots of nearby molecular material, forming
evaporating gaseous globules, or {\it EGGs} (Hester et al. 1996).

A new class of tadpole-shaped objects was recently
discovered near the Cygnus OB2 association (Sahai, Morris \& Claussen 2012\,[SMC12], Wright et al. 2012\,[Wetal12]) --  
the most massive young association within 2 kpc of the Sun, located inside
the Cygnus-X giant molecular cloud complex where considerable star formation is ongoing.
The strong morphological similarity between these objects and proplyds led Wetal12 to conclude that these objects
are ``a unique class of  photoevaporating, partially-embedded, young stellar objects" and unlikely to
be EGGs, from a comparison of the large fraction (about 70\,\%) of these objects
that appear to contain one or more young stars with the much lower value for M16 (15\%).
However, SMC12 concluded that Wetal12's argument is not compelling, as several unconsidered variables,
including density and mass of parent cloud, and strength of radiation field and winds, can affect
this fraction. In addition, the sizes of these objects (20,000--110,000\,AU) are
huge compared to the Orion proplyds (40--350\,AU: Henney \& O'Dell 1999).
Recognizing the importance of these issues, SMC12 carried out molecular-line observations that reveal
dense molecular cores associated with two of these objects (the Tadpole and the Goldfish), with total masses
of cold ($\sim10-15$\,K) molecular gas exceeding 1--2\,\ms, strongly favoring the
EGG hypothesis. These data also 
conclusively rule out the proplyd hypothesis since a proplyd is highly unlikely to harbor such a 
substantial mass of molecular gas. To date, no dense molecular medium has yet been found towards any known proplyd, and 
the low Orion proplyd disk masses (0.003--0.07)\,\ms~(Mann \& Williams 2010), derived from continuum observations of dust emission, 
imply even smaller masses for the circumstellar material being evaporated from them. Finally, the evaporated material is likely to be primarily atomic, not molecular.

SMC12 concluded that these tadpole-shaped objects are
dense, star-forming molecular cores that
originated in the Cygnus cloud and are now being photoevaporated by ultraviolet radiation from the
Cyg\,OB2\,No.\,8 cluster located $\sim$10\,pc to their North-West, and shaped by
ram pressure of strong, nearby wind sources. Continuing observations of $^{13}$CO and $^{12}$CO lines, 
as well as high-density tracers such as the J=3-2 lines
of HCO$^+$, HCN, HNC, and N$_2$H$^+$ in a sample of about 20 such objects have strongly supported the SMC12 hypothesis (Sahai et al. 2012b). 

The discovery of these free-floating Evaporating Gas Globules (hereafter frEGGs) raises a fundamental question for star and protoplanetary
disk formation in massive SFRs: what fraction of the previously classified proplyd objects are really EGGs? 
The distinction between frEGGs and proplyds is quite important, because in the former case, star formation and protostellar accretion are still very much under way, whereas proplyd disks are in much later stages in which accretion no longer has a strong effect on the protostellar evolution, although planet formation may still be in progress.
We have therefore begun a study of the molecular gas content of objects previously classified as proplyds with the goal of determining 
their true nature. Smith, Bally \& Morse (2003: SBM03) found many objects in the Carina SFR from ground-based optical emission-line images that closely resemble the proplyds found in Orion, and concluded that the former were proplyds as well  
in spite of their significantly larger (factor $\sim5$) sizes. In this Letter, we report single-dish molecular line observations of one such representative object 
from the set of 12 reported in SBM03, 104632.9-600354 (hereafter Carina-frEGG1), and show that our data rule out the proplyd hypothesis for this object and support its frEGG nature.

\section{Observations \& Results}
We observed Carina-frEGG1 (Fig.\,\ref{irachst}) in the CO J=3--2, 4--3, 6--5 and 7--6 lines, the $^{13}$CO J=3--2 and the HCN and HCO$^{+}$ J=4--3 lines, and the continuum at 350 and 870\micron~using the Atacama Pathfinder Experiment (G\"{u}sten et al. 2006) 
12-m telescope\footnote{APEX is a collaboration between the Max-Planck-Institut fuer Radioastronomie, the European Southern Observatory, and the Onsala Space Observatory} during June 23--26, 2012. Pointing was 
checked on nearby $\eta$\,Car, and found to be generally accurate within 3\arcsec.  

The dual-color receiver FLASH+ (Heyminck et al. 2006) with wideband (4\,GHz width) sideband-separating (2SB) SIS mixers was used to 
map the CO J=4--3 and 3--2 lines simultaneously in a regular raster 
with 10\arcsec~increments. Two XFFT spectrometers, each with a bandwidth of 2.5\,GHz, but with $64\times10^3$ ($32\times10^3$) channels for the low (high frequency) bands, were used to process a total IF band of 4\,GHz (with 1\,GHz of overlap).
The weaker $^{13}$CO(3--2), HCN and HCO$^{+}$(4--3) lines were observed towards the nominal center position only.

The dual-color heterodyne array receiver CHAMP+ (Kasemann et al. 2006), providing $2\times7$ beams, was used to map the CO J=6--5 and 7--6 lines simultaneously, over a $40\arcsec\times40\arcsec$ region that was oversampled with 4\arcsec~spacing (with all 7 beams of each sub-array covering a given grid position repeatedly). Two FFT spectrometers, each with a bandwidth of 1.5\,GHz and 1024 channels, were connected to the individual beams, processing a total IF band of 2.6\,GHz (with 400\,MHz overlap). The observations took place during excellent weather conditions with zenith precipitable water vapor, $PWV$, of 0.7\,(0.4)\,mm in the low (high) frequency band.

The spectra were taken with position-switching against an absolute reference position. 
Calibration was performed regularly every 10--15\,min with a cold liquid nitrogen (LN2) load and an ambient temperature load. The data were processed with the APEX real-time calibration software (Muders et al. 2006). Beam sizes (FWHM: full-width at half-maxiumum) and main beam coupling efficiencies are 
given in Table\,\ref{mmdata}.

We obtained continuum observations at 350 (870)\,\micron~using the bolometer array receivers LABOCA (SABOCA) (Siringo et al. 2009, 2010), with a spatial resolution of $7.8{''}$ ($19{''}$) on 2012 Sep 10 (Sep 11) and $PWV\sim0.5\,(0.3-0.4)$\,mm. Data reduction was done with the BOA software (Schuller et al. 2009) following standard procedures, including iterative source modeling.

\begin{figure}[!h]
\resizebox{0.7\textwidth}{!}{\includegraphics{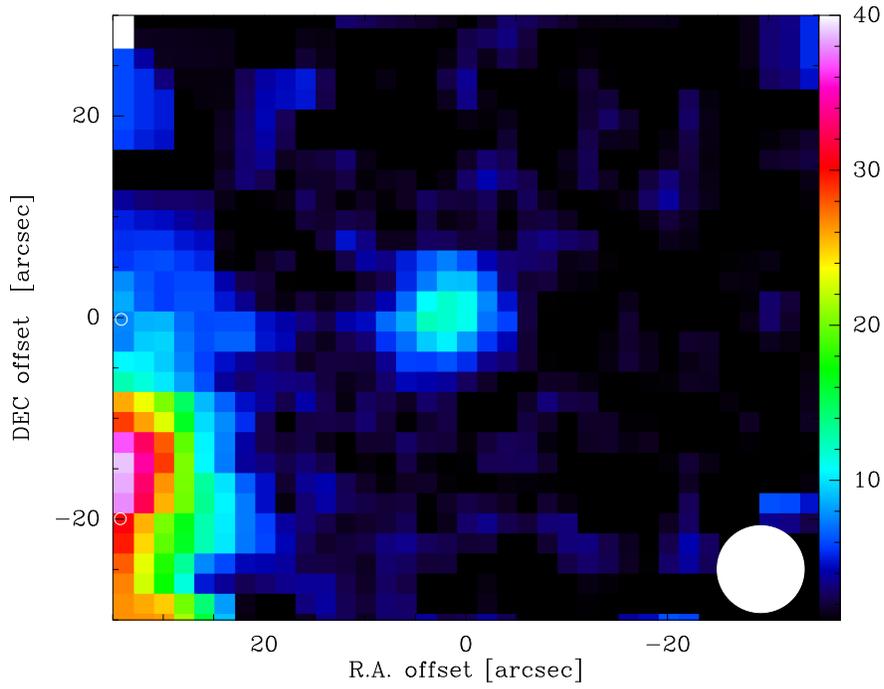}}
\caption{Map of Carina-frEGG1 in CO J=6--5 emission integrated over the central 5\,\kms~of the line profile, obtained with the APEX-12m. White circle shows the 8.7\arcsec~beam-size (FWHM).}
\label{co65map}
\end{figure}

\begin{figure}[!h]
\resizebox{1.0\textwidth}{!}{\includegraphics{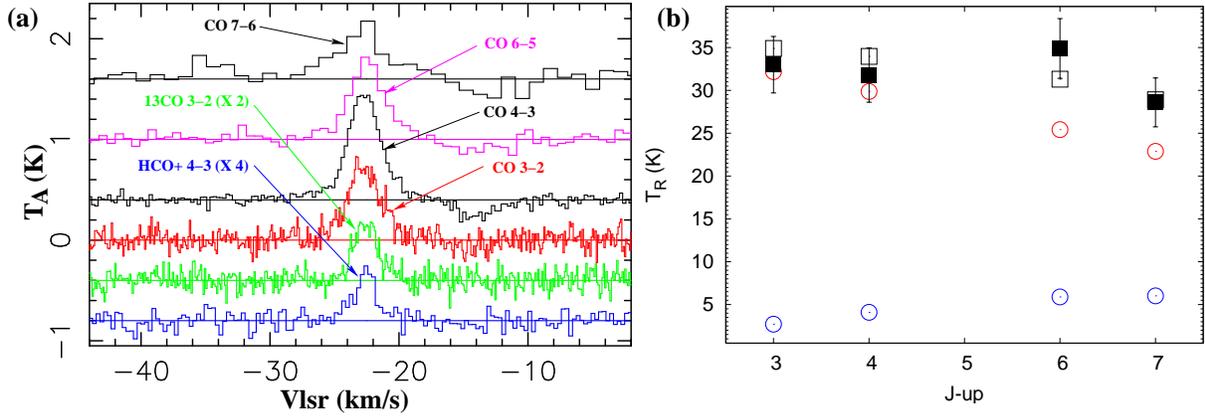}}
\caption{(a) Molecular line emission from Carina-frEGG1 observed with the
APEX-12m. The line profiles have been rescaled and shifted vertically for clarity. Weak absorption features in the CO J=4--3, 6--5 and 7--6 linse at $V_{lsr}\sim -11$ to 
$-16$\,\kms~are due to emission in the reference position used for the position-switching. Intensities have not been corrected for main-beam efficiencies. (b) Observed radiation temperatures of CO lines from Carina-frEGG1 ({\it filled black squares}) with 20\% error bars, and a two-component model fit ({\it open black squares}). The lower temperature,  optically-thick component at 40\,K ({\it red}) fits only the 3--2 and 4--3 intensities; a second, hotter ($\sim150$\,K), optically-thin component ({\it blue}) is needed to fit the low and high-J line intensities simultaneously.
}
\label{spec}
\end{figure}

The CO maps show the presence of a compact molecular globule (Fig.\,\ref{co65map}), that appears unresolved at our highest 
angular resolution of 7.7\,\arcsec~in the CO J=7--6 line.  The line profiles 
(Fig.\,\ref{spec}a) show a central core and weak wings, extending to about $\pm6$\,\kms~from the line center as seen in our highest S/N profile (J=4--3), and suggesting the presence of an outflow. Gaussian fits to the line profiles result in a width (FWHM) of about 3.1\,\kms~in the 3--2 and 4--3 lines, and 3.5 and 4.1\,\kms~ in the 6--5 and 7--6 lines, respectively, implying an increase in the wing contribution relative to the core for the higher excitation lines. The J=4--3 and 3--2 maps show the presence of low-level extended background or foreground emission around the compact source seen in the J=6--5 and 7--6 lines, but the contribution is small ($\lesssim$10\%).

In order to determine relative line intensity ratios, we convolved the CO 6--5 and 7--6 maps to the 4--3 map resolution, $13.1{''}$ (we chose not to convolve these maps to the larger 3--2 beamsize because of possible contamination from the extended cloud to the East).
Since the source is unresolved, the source's average brightness temperature in each line, $T_R$, 
depends on its emitting area. 
Conservatively assuming that all CO lines come from the dark region within the (ionized) bright periphery seen in the optical image, 
that has an area of 11.7\,arcsec$^2$, we derive the radiation 
temperature, $T_R$, for each of the observed CO lines (Table\,1). Using the online tool for the non-LTE radiative transfer code, RADEX (Van der Taak et al. 2007) to model these, we find that the 
4--3 and 3--2 data can be fitted with a column density N(H$_2$)$\sim10^{22}$ cm$^{-2}$ and kinetic temperature
$T_{k}\sim 40$\,K (assuming a CO/H$_2$ abundance ratio $f_{CO}\sim10^{-4}$), but the 6--5 and 7--6 line intensities predicted by this one-temperature model are weaker than observed 
(Fig.\,\ref{spec}b), even with a model density high enough (\dens$\sim 10^6$cm$^{-3}$) to thermalize these lines, implying the presence of 
gas with $T_{k}>>40$\,K in the globule. The ``excess" emission in the 6--5 and 7--6 lines requires a hotter ($\gtrsim150$\,K) optically-thin component  assuming it has the same emitting area as the cooler component. The observed CO 6--5, 7--6 and HCO$^+$ 4--3 line intensities  
imply the presence of high gas densities, $\gtrsim\,5\times10^5$\,cm$^{-3}$.

We derived the molecular mass in Carina-frEGG1 from its $^{13}$CO J=3--2~flux, using Eqn. (5) of Thi et al (2001), who used it to estimate protoplanetary disk gas masses from observations with the JCMT 15-m telescope (14\arcsec~beam). We modified this equation to account for the APEX-12m's larger beam-size. 
We estimate a ``beam-averaged" $^{13}$CO J=3--2 optical depth, $\tau _{13}=0.56$ from the $^{12}$CO/$^{13}$CO line intensity ratio, 
assuming \tex($^{13}$CO)=\tex($^{12}$CO)=40\,K. Assuming a $^{13}$CO/H$_2$ abundance ratio $f_{13CO}=10^{-6}$ and distance 2.3\,kpc (e.g., Smith et al. 2004), we find a molecular mass of $M_{mol}=0.35$\,\ms.

A $196\pm65$\,mJy continuum source  was detected at 350\,\micron, and an upper limit (40\,mJy, $1\sigma$) obtained at 870\,\micron. Assuming the 350\,\micron~flux is due to optically-thin, thermal emission from dust at the same temperature as we derive above for the bulk of the gas (i.e., $T_d=T_{k}\sim40$\,K), a gas-to-dust ratio of 100, and a dust opacity $\kappa(350\,\micron)=10$\,cm$^2$\,g$^{-1}$ (e.g., Miettinen et al. 2012), we derive a total mass of $\sim0.1$\,\ms. The discrepancy between this value and the mass derived from CO emission is not significant, considering the uncertainties in the adopted values of $T_d$, $\kappa(350\,\micron)$ and $f_{13CO}$.

\section{Additional Multi-wavelength Data}
\begin{figure}[!h]
\resizebox{1.0\textwidth}{!}{\includegraphics{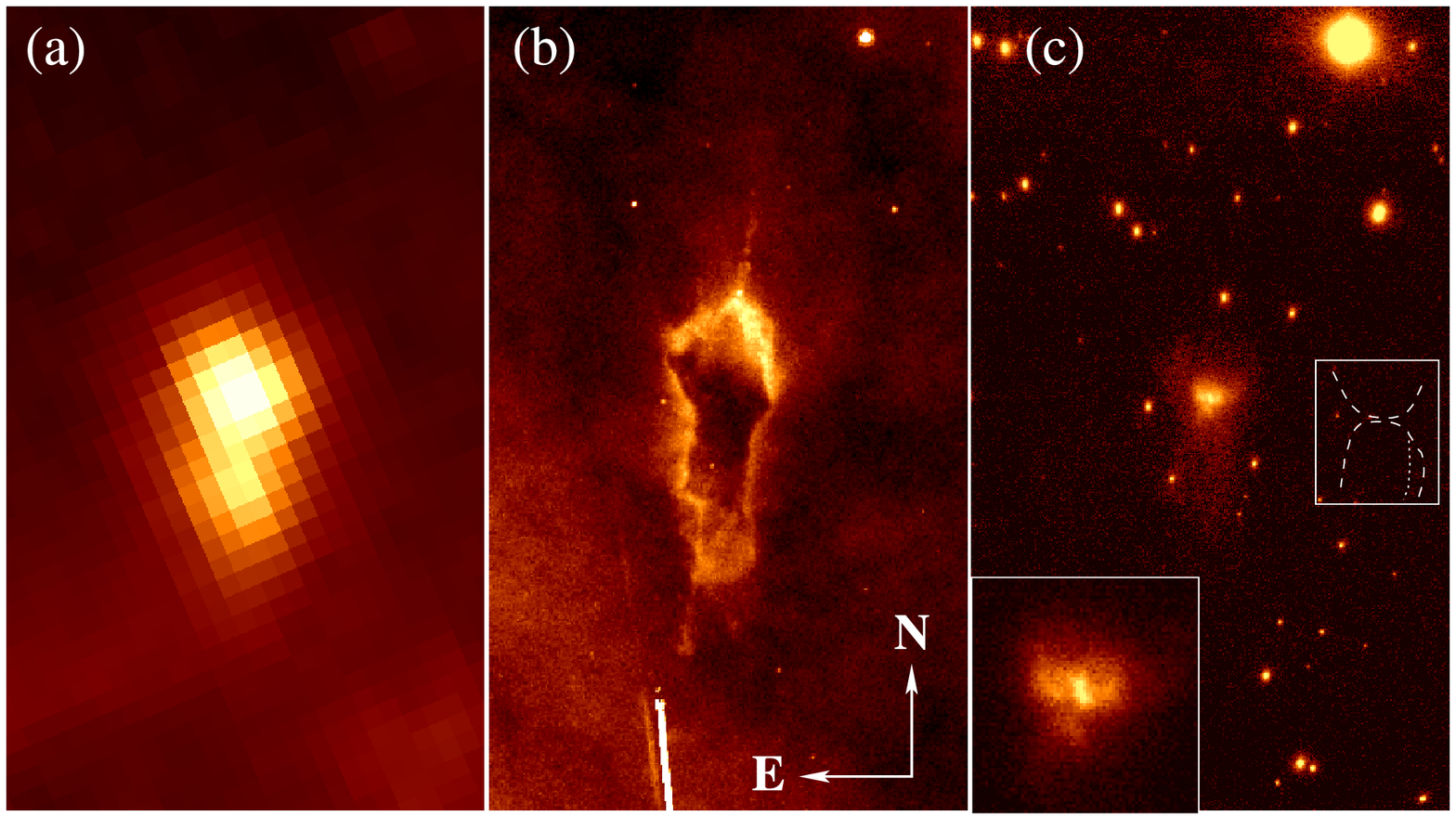}}
\caption{Images of Carina-frEGG1 at different wavelengths (a) Spitzer IRAC 8\micron, 
(b) HST F658N narrow-band (F658N filter: H$\alpha$+[NII]), and (c) VLT/NACO Ks (2\micron). 
The panels are $14.6\arcsec\times24.4\arcsec$, with North up (as shown). Inset at right-middle shows a schematic of the hourglass-shaped central nebula (two possible curves are shown for the southern lobe's poorly-defined western periphery); inset at bottom-left corner shows an enlarged view ($1.89\arcsec\times1.96\arcsec$) of the central source. 
}
\label{irachst}
\end{figure}

We have compiled multi-wavelength data on Carina-frEGG1 using optical (HST), near-IR (2MASS and ESO) and mid-IR (Spitzer) data archives. Carina-frEGG1 was imaged with ACS in H$\alpha$ as part of a large-scale survey of the Carina SFR (Smith, Bally \& Walborn 2010), and shows a dark tadpole-shaped globule (of length $\sim$19,000\,AU) with a bright periphery. 
The HST H$\alpha$ (+[NII]) image of Carina-frEGG1 shows a narrow, highly collimated jet (Fig.\,\ref{irachst}b) -- Smith et al. (2010) discuss this jet (HH1006) and infer that the putative central star related to the jet is not detected, presumably as a result of being deeply embedded within dust in the object. 

The 2MASS images show a faint red compact source in 
the J, H, and Ks bands. Using two nearby field stars to register the ACS and 2MASS Ks images, we find that the source is located at 
$R.A.=10^h46^m32^s.97, Dec=-60\arcdeg03^m53^s.5$, roughly in the middle of the dark waist. We also retrieved archival Ks images towards Carina-frEGG1 taken with the NACO+CONICA instrument on the ESO-VLT-U4 on 10 Jan 2007, with a total field-of-view $27.8\arcsec\times27.8\arcsec$. Using standard IRAF packages, we generated a flat-field from these images using median-averaging and applied it to each of the 17~30-sec dithered exposures, which were then registered to a common reference frame and averaged. The angular resolution in the final image is about $0\farcs1$ as measured from radial intensity cuts of several neighbouring field stars.  We find an extended faint hourglass-shaped nebula with a bright central region (Fig.\,\ref{irachst}c) with its symmetry axis aligned with that of the jet. The northern lobe is much brighter than the southern one, implying that the hourglass structure is tilted such that its northern lobe is closer to us. In the central region of the northern lobe, which has a wide-U shape with its apex located at the center of the hourglass (inset, Fig.\,\ref{irachst}c), the brightest feature is a small spur (of length $\sim$50\,AU) oriented at a position angle, 
$PA\sim12\arcdeg$. The spur is located slightly west of the jet axis and is not aligned with the latter (which is oriented at $PA\sim-10\arcdeg$). 

The Spitzer IRAC images (3.6--8\,\micron) show an elongated morphology similar to Carina-frEGG1's optical shape (but at much lower angular resolution). As in the Ks image, the shape is bipolar, with the northern half significantly brighter than the southern one. No central star can be seen in the Ks image or the IRAC images, implying that the former is deeply embedded within a dusty disk with very high line-of-sight extinction. 
In longer-wavelength Spitzer images 
(i.e., $\ge24$\micron), no isolated source can be detected at the location of Carina-frEGG1 due to bright extended emission from foreground or background clouds in its vicinity. 

\begin{figure}[!h]
\resizebox{0.7\textwidth}{!}{\includegraphics{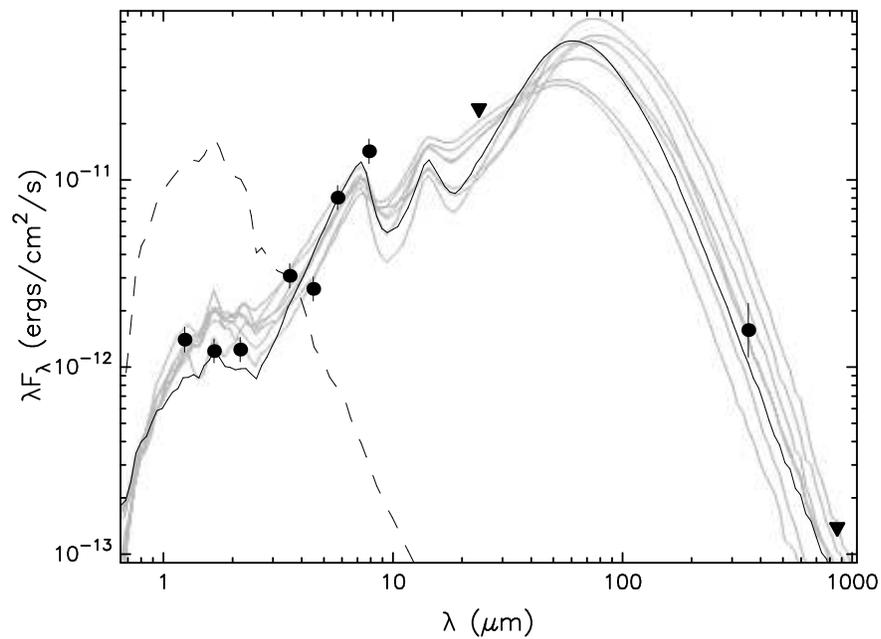}}
\caption{Spectral energy distribution of Carina-frEGG1. Errors in the photometry are taken to be $\pm$15\% for data from 1.24\,\micron~to 8\,\micron, and $\pm$30\% at 350\,\micron; black triangles show upper limits. The black curve shows the best-fit YSO model assuming ISM extinction $A_V=1.5$ and distance 2.3\,kpc, and the gray lines show subsequent good fits.  The dashed line shows the stellar photosphere from the best-fit model in the absence of circumstellar dust (but including ISM extinction).
}
\label{sed}
\end{figure}

We have determined the SED of Carina-frEGG1 from 1.25 to 870\,\micron. We carried out aperture photometry on the J and H-band 2MASS images (since the 2MASS catalog gives only a Ks magnitude). Aperture photometry for IRAC (channel 1 to 4) and MIPS (channel 1) is provided by 
Povich et al. (2011: source PCYC\,1173 in their catalog). Since no compact source can be seen in the MIPS 24\,\micron~image at Carina-frEGG1's location, we treat Povich et al.'s measured 24\,\micron~magnitude of 3 as  
an upper limit. The SED 
(Fig.\,\ref{sed}) continues to rise towards 350\,\micron~-- the longest wavelength for which we have a detection -- indicating a substantial amount of emission from cool circumstellar dust in the far-IR range. For the source luminosity, we integrate the SED up 
to 350\,\micron~to estimate $L\gtrsim8\,\ls$; a rough upper limit of $18\,\ls$~is provided by assuming flux values equal to the upper limits at 24 and 870\,\micron. The observed fluxes have been corrected using an interstellar extinction $A_V=1.5$, that is intermediate between the value (1.78) given by Smith et al. (2004) and that (1.3) computed from a numerical algorithm provided by Hakkila et al. (1997), which estimates the $A_V$ of a galactic source from its longitude, latitude and distance. 

We modeled the SED using an online tool that applies least-squares fitting to find the best models from a large set of 
pre-computed YSO models having accretion disks and infall envelopes with biconical outflow cavities (Robitaille et al.\,2007). The ten best-fit models (Fig.\,\ref{sed}) have relatively low stellar effective temperatures (2660--3800\,K), stellar masses (0.1--0.65\,\ms) and total luminosities (stellar\,+\,disk-accretion: 6--10\,\ls) and small ages (1100--3000 yr). These results support our inference of a 
deeply-embedded YSO in Carina-frEGG1, but must be treated with some caution. Although the inferred inclination of the outflow to the line-of-sight, $i=18$\arcdeg, together with the large circumstellar extinction along the disk's mid-plane ($A_V\gtrsim100$), is qualitatively consistent with the strong asymmetry in the brightness of the bipolar lobes in the Ks image, the projected opening angle of the cavity, 
$\theta_{proj}=tan^{-1}(tan\,\theta_c/sin\,i)=23$\arcdeg~($\theta_c\sim7$\arcdeg, is the intrinsic opening angle in these models) 
appears smaller than observed. 

\section{Discussion}
Our detection of a substantial mass of molecular gas in Carina-frEGG1 shows that it is an frEGG, not a proplyd (as classified previously).
Since the discovery of proplyds in Orion, many studies have reported the finding of  
proplyds in other SFRs with massive stars (e.g., Balog et al. 2006, Koenig et al. 2008, Brandner et al. 2000). 
However, in light of this study and that in SMC12, it is likely that the true nature of many or all of these objects has been misunderstood, and that some (or even all) of the 
previously classified proplyds in Carina, especially those which are significantly larger in size than the Orion proplyds,
are really frEGGs. 

Both proplyds and EGGs are unique probes of the effects of
the harsh UV radiation and strong stellar winds from massive stars on the formation of lower-mass stars in their vicinity, thus it
is of fundamental importance to be able to distinguish between these two classes. EGGs are most likely the surviving high
density concentrations in a cloud as the ionization front sweeps through it. The formation of a protostar in the EGG 
may be induced by the compression of the latter by the high-pressure
environment of the ambient HII region, supplemented by the even higher pressure at the
high-density ionization front at the EGG surface. 
Photoevaporation sculpts out and exposes an EGG while the stellar object(s) in it is (are) still accreting mass, ultimately freezing the protostellar mass distribution at a relatively early stage in its evolution (e.g., Hester et al. l996). FrEGGs like Carina-frEGG1 (where the presence of collimated jets, central hourglass structure, and low luminosity, all imply the presence of a jet-driving, young, low-mass star) or the Tadpole (SMC12) represent the early formation phase of low or intermediate-mass stars within a massive SFR. Proplyds may represent the 
endpoint of evolution of such frEGGs, or may form coevally with the OB stars in these SFRs. 
Whether other objects like Carina-frEGG1 are proplyds or EGGs has major implications for the 
star-formation rate and the IMF in the massive SFRs where they are found. Outstanding questions that future research should address are: (i) do proplyds represent the evolutionary endpoint of frEGGS and (ii) is the formation of stars that may form in frEGGs induced by the surrounding HII region's compression or were such stars already in the process of formation in the relatively dense cloud cores that are susceptible to becoming frEGGS?   
\section{Acknowledgments}  We thank F. Schuller, APEX, for his help with the LABOCA and SABOCA observations and data reduction. RS's contribution to the research described here was carried out at JPL, California Institute of Technology, under a contract with NASA and partially funded through the internal Research and Technology Development program.

\begin{table}
   \caption{Molecular Line Observations of Carina-frEGG1}
   \label{table1}
  \begin{tabular}{l c c c c c c c }
\hline\noalign{\smallskip}
Line & \tmb\tablenotemark{a} & $T_R$\tablenotemark{b}   &\vlsr  & Line\,Width & Line\,Flux & $\eta _{mb}$ & $\theta _b$\\
     & (K)          & (K)   & (km\,s$^{-1}$) & (km\,s$^{-1}$) & (K\,km\,s$^{-1}$) & & (${''}$) \\
\hline\noalign{\smallskip}
CO(3-2)       & 1.1  & 33.0 & -22.7 &  3.2  & 3.5   & 0.69 & 17.8 \\
$^{13}$CO(3-2)& 0.4  & 13.4 & -22.6 &  2.2  & 0.96  & 0.69  & 18.6 \\
CO(4-3)       & 1.9  & 31.8 & -22.7 & 3.1  & 5.4   & 0.60  & 13.1 \\
CO(6-5)       & 2.1\tablenotemark{c}  & 34.9 & -22.6 &  3.5  & 7.3\tablenotemark{c}  & 0.38 & 8.7 \\
CO(7-6)       & 1.7\tablenotemark{c}  & 28.6 & -22.3 &  4.1  & 8.1\tablenotemark{c}  & 0.31 & 7.7 \\
HCO$^{+}$(4-3)& 0.17 &  ... & -22.6 &  2.0  & 0.37  & 0.69 & 17.3 \\
HCN(4-3)      & 0.11 &  ... & -22.8 &  1.8  & 0.21  & 0.69 & 17.4 \\ 
\hline\noalign{\smallskip}
  \end{tabular}
\label{mmdata}
\tablenotetext{a}{Main-beam line-intensities}
\tablenotetext{b}{Source brightness temperature derived by applying beam-dilution corrections}
\tablenotetext{c}{Derived from peak position in map convolved to $13.1{''}$ resolution}
\end{table}

\end{document}